\documentclass[aps,twocolumn,showkeys,showpacs,preprintnumbers,prd,superscriptaddress,nofootinbib,10pt]{revtex4-1}
\usepackage{graphicx,epsf,bm,amsmath,amsfonts,amssymb,epstopdf,natbib,hyperref,xcolor,appendix,verbatim,multirow,diagbox,booktabs,siunitx}
\usepackage[normalem]{ulem}
\usepackage{array}
\usepackage{diagbox}
\usepackage{makecell}
\usepackage[makeroom]{cancel}
\usepackage{fontawesome}
\usepackage{hyperref}
\hypersetup{colorlinks=true,urlcolor=blue,citecolor=blue,linkcolor=blue,menucolor=blue,anchorcolor=blue,filecolor=blue}
\renewcommand{\arraystretch}{1.5}
\newcolumntype{?}{!{\vrule width 3pt}}
\newcommand{\thickhline}{\noalign{\hrule height 1.5pt}}

\usepackage{graphicx} 
\usepackage{array}    
\usepackage{makecell} 

\newcolumntype{?}{!{\vrule width 1.5pt}}

\date{\today}

\makeatletter
\newcommand{\github}[1]{
   \href{#1}{\faGithubSquare}}
\makeatother
\newcommand{\fmarki}{\ensuremath{\alpha}}
\newcommand{\fmarkii}{\ensuremath{\beta}}
\newcommand{\fmarkiii}{\ensuremath{\gamma}}
\newcommand{\fmarkiv}{\ensuremath{\delta}}
\newcommand{\fmarkv}{\ensuremath{\epsilon}}
                
\def\@fnsymbol#1{{\ifcase#1\or \fmarki\or \fmarkii\or \fmarkiii\or \fmarkiv\or \fmarkv\or \else\@ctrerr\fi}}
\makeatother
\definecolor{darkred}{rgb}{0.8, 0.0, 0.0}
\definecolor{darkgreen}{rgb}{0, 0.8, 0.0}

\definecolor{valecol}{rgb}{0,0.5, 1.}

\begin{document}

\title{Geometric Constraints on the Pre-Recombination Expansion History\\ from the Hubble Tension}

\author{Davide Pedrotti}
\email{davide.pedrotti-1@unitn.it}
\affiliation{Department of Physics, University of Trento, Via Sommarive 14, 38123 Povo (TN), Italy}
\affiliation{Trento Institute for Fundamental Physics and Applications (TIFPA)-INFN, Via Sommarive 14, 38123 Povo (TN), Italy}

\begin{abstract}
\noindent
I perform a model-independent reconstruction of the background pre-recombination expansion history of the Universe. I find that purely early-time resolutions to the Hubble tension, satisfying the geometric CMB constraints, exist at the background level. This class of solutions requires a smooth transition around matter-radiation equality, characterized by a $\simeq 15\%$ expansion rate enhancement prior to recombination. This result serves as a blueprint for future model-building approaches, providing a background stress-test for Hubble tension proposals.
\end{abstract}

\maketitle
\textbf{\textit{Introduction.}} --- The persistent discrepancy between early- and late-time determinations of the Hubble constant has evolved into a definitive crisis for modern cosmology \cite{Verde:2019ivm, DiValentino:2021izs,DiValentino:2020vvd,Perivolaropoulos:2021jda,Shah:2021onj,Abdalla:2022yfr,DiValentino:2022fjm, Hu:2023jqc, Verde:2023lmm,CosmoVerseNetwork:2025alb}. Recent local distance ladder measurements establish $H_0 = (73.50 \pm 0.81)$ km/s/Mpc \cite{H0DN:2025lyy}, disagreeing at $>7 \sigma$ with the value inferred from the high-redshift universe assuming the standard $\Lambda$CDM model, $H_0 = (67.24 \pm 0.35)$ km/s/Mpc \cite{SPT-3G:2025bzu}. This ``Hubble tension" provides a compelling case for new physics \cite{Anchordoqui:2015lqa, DiValentino:2016hlg,Benetti:2017juy,Kumar:2017bpv,Graef:2018fzu, Yang:2018euj,Guo:2018ans,Kreisch:2019yzn,Vagnozzi:2019ezj,DiValentino:2019ffd,DiValentino:2019jae,Krishnan:2020obg,Alestas:2020mvb,RoyChoudhury:2020dmd,Brinckmann:2020bcn,Gao:2021xnk,Marra:2021fvf,Dainotti:2021pqg,Krishnan:2021dyb,Cyr-Racine:2021oal,Anchordoqui:2021gji,Schoneberg:2021qvd,Akarsu:2021fol,Banerjee:2020xcn,deSa:2022hsh,Akarsu:2022typ,Bernui:2023byc,Gomez-Valent:2023hov,Ruchika:2023ugh,Adil:2023exv,Frion:2023xwq,Gomez-Valent:2023uof,Teixeira:2024qmw,Akarsu:2024qiq,Akarsu:2024qsi,Giare:2024ytc,Giare:2024smz,Escamilla:2024xmz,Nozari:2024wir,RoyChoudhury:2024wri,Mirpoorian:2024fka, Pedrotti:2025ccw, Zhang:2025dwu,Poulin:2024ken,Li:2025owk,Lee:2025yah,Teixeira:2025czm,Wang:2025dzn,Zhang:2025dwu,Hogas:2025mii,Pantos:2026rpe}. Central to understanding this tension is the angular scale of the sound horizon at recombination, a fundamental quantity measured in the Cosmic Microwave Background (CMB) with exquisite precision \cite{Planck:2018vyg}:
\begin{equation}
    \theta_s^\star = \frac{r_s^\star}{D_A^\star} = \frac{H_0 r_s^\star}{\int_0^{z_\star} dz/E(z)}\ , \label{eq:theta_s}
\end{equation}
where $r_s^\star$ is the sound horizon at recombination $z_\star$, $D_A^\star$ is the angular diameter distance to the last scattering surface, and $E(z)$ is the unnormalized expansion history of the Universe $E(z) \equiv H(z)/H_0$. To reconcile a high local $H_0$ with a tightly constrained $\theta_s^\star$, theoretical efforts divide into two broad categories. \textit{Late}-time proposals attempt to modify the unnormalized expansion history $E(z)$ at low redshifts. However, these are severely disfavored by no-go theorems based on low-redshift probes like BAO and Type Ia Supernovae \cite{Bernal:2016gxb, Addison:2017fdm, Lemos:2018smw, Aylor:2018drw, Knox:2019rjx,Cai:2021weh, Schoneberg:2019wmt, Arendse:2019hev,Efstathiou:2021ocp,Keeley:2022ojz,Zhou:2025kws,Huang:2024erq,Bansal:2026axl}, leaving little room for late-time deviations from $\Lambda$CDM. Consequently, focus has shifted to \textit{early}-universe mechanisms, such as Early Dark Energy \cite{Karwal:2016vyq,Poulin:2018cxd,Smith:2019ihp, Niedermann:2019olb, Niedermann:2020dwg,Ye:2020btb,Smith:2020rxx, Herold:2021ksg, Karwal:2021vpk,Kamionkowski:2022pkx,Poulin:2023lkg,McDonough:2023qcu} or extra relativistic species \cite{Mortsell:2018mfj,Escudero:2019gzq, Escudero:2019gvw,Pandey:2019plg,Archidiacono:2019wdp,Blinov:2020hmc,Nakai:2020oit,Aloni:2021eaq,Berghaus:2022cwf,Schoneberg:2022grr}, that aim to reduce $r_s^\star$.

Crucially, as highlighted in \cite{Knox:2019rjx}, most early-time models fall short to fully restore concordance because they struggle to reduce $r_s^\star$ without altering three fundamental CMB scales: the angular scales projected to the last scattering surface of the comoving particle horizon at matter-radiation equality ($\theta_\text{eq}$), the comoving sound horizon at recombination ($\theta_s^\star$), and the photon diffusion damping scale ($\theta_d^\star$) \cite{Hu:2001bc, Hou:2011ec}. Notice, however, that as stressed in \cite{Adil:2022hkj}, the fit to the three abovementioned angular scales provides a \textit{necessary} but not \textit{sufficient} condition to an overall good fit to the full CMB power spectrum.

In this \textit{Letter}, I adopt a strictly model-agnostic approach to early-time resolutions to the Hubble tension. Rather than positing a specific physical mechanism, I perform a non-parametric reconstruction of the pre-recombination expansion history to pinpoint the exact background requirements needed to resolve the tension while preserving the CMB angular scales. While reconstruction methods have been widely applied to the late-time Universe and dark energy \cite{Sahni:2006pa,Shafieloo:2007cs,Seikel:2012uu,Holsclaw:2010sk,Mamon:2016dlv,Jiang:2024xnu,Velazquez:2024aya,Sabogal:2024qxs,Gonzalez-Fuentes:2025lei,Akarsu:2026anp,Verdugo:2026oqd}, as well as to various early-dark energy parametrizations \cite{Hojjati:2013oya,Linder:2010ks,Samsing:2012qx,Moss:2021obd,Kou:2024rvn,Bella:2026zuk}, this work represents the \textit{first model-independent reconstruction of the pre-recombination expansion history guided specifically by CMB geometric constraints and the Hubble tension}, to the author’s knowledge. Notably, this framework focuses on the background evolution and does not capture proposals that modify the thermal history or the sound horizon scale $r_s^\star$ through non-standard recombination physics \cite{Hart:2017ndk,Hart:2019dxi,Sekiguchi:2020teg,Jedamzik:2020krr,Galli:2021mxk,Lee:2022gzh,Lynch:2024hzh,Lynch:2024gmp}.

I find that purely early-time resolution to the Hubble tension, satisfying the
geometric CMB constraints, exist at the background level. This class of solutions intrinsically requires a background expansion history characterized by a smooth transition around matter-radiation equality with a pronounced ($\simeq 15\%$) departure from the fiducial $\Lambda$CDM immediately before recombination. While future dedicated studies must assess the full perturbation-level compatibility of this feature with the CMB power spectrum, this severe geometric rigidity highlights the challenge of fully pre-recombination solutions. On the other hand, because the \textit{shape} of this mathematical solution is independent of a specific amplitude, any successful early-universe mechanism must mimic this smooth transition. Ultimately, this rigid background profile serves as a fundamental \textit{blueprint for future model building}, both for fully pre-recombination proposals and for hybrid cosmologies, that distribute departures from $\Lambda$CDM across both the pre- and post-recombination eras \cite{Jedamzik:2020zmd,Khosravi:2021csn, Clark:2021hlo, Wang:2022jpo, Anchordoqui:2022gmw, Reeves:2022aoi,Yao:2023qve, daCosta:2023mow, Wang:2024dka, Vagnozzi:2023nrq,Toda:2024ncp,Gonzalez-Fuentes:2026rgu}. 

\textbf{\textit{Analysis setup.}} --- The no-go theorem for late-time solutions to the Hubble tension strongly suggests early-time modifications to $\Lambda$CDM. Specifically, based on Eq.~(\ref{eq:theta_s}) and the low- and high-redshift $H_0$ measurements reported above, a successful early-time modification must reduce the sound horizon scale at recombination by approximately $\delta r_s^\star/r_s^\star \simeq -7\%$, without altering the fit to the CMB power spectrum. This means that $\delta D_A^\star/D_A^\star$ must be reduced by roughly the same amount, to keep $\theta_s^\star$ from changing too much. Furthermore, to maintain the fit to all the three angular scales mentioned above, the ratios of their associated length scales must also not change by too much. I use these geometric constraints as conditions for a model-independent reconstruction of $\delta H / H_{\text{fid}} \equiv (H(z) - H_\text{fid}(z)) / H_\text{fid}(z)$, where $H_\text{fid}(z)$ is the Hubble parameter of the \textit{Planck 2018} bestfit flat-$\Lambda$CDM. In summary:
\begin{itemize}
    \item $\delta r_s^\star/r_s^\star\simeq -7\%$, 
    \item $\theta_s^\star /\theta_d^\star \sim \text{const} \Longrightarrow \delta r_d/r_d \simeq -7\%$, 
    \item $\theta_s^\star/\theta_{\text{eq}} \simeq \text{const} \Longrightarrow \delta r_\text{eq}/r_\text{eq} \simeq -7\%$.
\end{itemize}
The treatment of the sound horizon and damping scales at recombination is straightforward. By definition: 
\begin{equation}
    r_X = \int_{z_*}^{\infty} \frac{f_X(z)}{H(z)} dz, \qquad _X = s,d
    \label{eq:rX}
\end{equation}
where:
\begin{align}
        f_{r_s} &= \frac{c_s(z)}{H(z)}\ \text{,} \quad c_s(z) = \sqrt{\frac{1}{3(1+R)}} \label{eq:f_rs},\\
        f_{r_d} &= \frac{\mathcal{F}(z)}{H(z)}\ \text{,} \quad \mathcal{F}(z) = \pi\sqrt{\frac{\left(\dfrac{R^2}{1+R} + \dfrac89\right)}{6H(z)n_e\sigma_T(1+z)^2}}. \label{eq:F_rd}
\end{align}
Here, $c_s(z)$ is the primordial plasma sound speed, $R(z) \equiv 3\rho_b/4\rho_\gamma$, while $n_e$ and $\sigma_T$ denote the free electron number density and the Thomson scattering cross-section, respectively. The functional derivative of the two scales with respect to the expansion rate is
\begin{equation}
    \frac{\delta \ln r_X}{\delta \ln H}= -\frac{1}{r_X}\frac{f(z)}{H(z)} \equiv K_X(z),
\end{equation}
which leads to,
\begin{equation}
    \delta r_X/r_X = \int_{z_\star}^\infty dz K_X(z) \frac{\delta H}{H_\text{fid}}(z),
\end{equation}
which can finally be used to reconstruct $\delta H/H_{\text{fid}}$.

The treatment of the condition on $\theta_\text{eq}$ is far less trivial, as when components that are neither matter nor radiation are injected before recombination, the matter-radiation equality scale becomes poorly defined. However, starting from the definition of the particle horizon scale at matter-radiation equality: 
\begin{equation}
    r_\text{eq} \equiv 1/(a_\text{eq}H(z_\text{eq})) \simeq \frac{z_\text{eq}}{H(z_\text{eq})},
    \label{eq:mr_scale}
\end{equation}
and perturbing $H(z)$ around the fiducial, I find:
\begin{equation}
    \delta r_\text{eq}/r_\text{eq} \simeq -\frac{\delta H}{H_\text{fid}}\Bigg|_{z=z_\text{eq}} + \frac{\delta z_\text{eq}}{z_\text{eq}}.
    \label{eq:condition_theta_eq}
\end{equation}
By requiring the redshift of matter-radiation equality not to change by too much to preserve the CMB peak structure, $\delta z_\text{eq}/z_\text{eq} \simeq 0$, Eq.~(\ref{eq:condition_theta_eq}) provides a \textit{necessary} condition that relates the amplitude of $\delta H / H_\text{fid}$ at the fiducial matter-radiation equality with $\delta r_\text{eq}/r_\text{eq}$. 

Finally, I reconstruct $\delta H/H_{\text{fid}}$ by numerically solving the following system of integral equations and the condition on $\delta H/H_\text{fid}$ at equality, which reflect the requirements listed above:
\begin{align}
    &\delta r_s^\star/r_s^\star\simeq -0.07 \simeq \int_{z_\star}^{\infty} dz K_s(z)\frac{\delta H}{H_\text{fid}}(z), \label{eq:deltars} \\
    &\delta r_d^\star/r_d^\star \simeq -0.07 \simeq \int_{z_\star}^{\infty} dz K_d(z)\frac{\delta H}{H_\text{fid}}(z), \label{eq:deltard} \\
    & \delta r_\text{eq}/r_\text{eq} \simeq -0.07 \simeq -\frac{\delta H}{H_\text{fid}}\Bigg|_{z=z_\text{eq}}.  \label{eq:deltareq}
\end{align}
Because the sensitivity kernel for the sound horizon $K_s(z)$ features a long, slowly decaying tail, deep into the radiation-dominated era, the $H_0$ tension could theoretically receive contributions from redshifts up to $z \sim 10^5$. However, since the kernel amplitude in this regime is exceedingly small (e.g., contributing only $\sim 3\%$ to the total integral between $z=20000$ and $30000$), generating any meaningful background shift here would require exotically large oscillations in $\delta H(z)/H(z)_\text{fid}$. By imposing the physical prior that the early universe must smoothly converge to standard $\Lambda$CDM without extreme deviations, the required modifications to $H(z)$ are naturally driven to lower redshifts ($z < 10000$) where the kernel carries sufficient support. I implement this by simply restricting my integration domain to $z_\text{max} \equiv 20000$. Additionally, I introduce a penalty term to suppress undesired, high-frequency oscillations, which are otherwise prone to emerge at high redshifts ($z \gtrsim 10000$) where the $r_s^\star$ and $r_d^\star$ kernels vanish \cite{2022MNRAS.514.6203T}:
\begin{equation}
    \mathcal{L}_{\text{penalty}} = -\lambda\int_{z_\star}^{z_{\text{max}}} dz \left(\frac{d^2}{dz^2}\left[\frac{\delta H}{H}\right]\right)^2.
    \label{eq:pen_like}
\end{equation}
The parameter $\lambda$ was chosen in such a way that $\mathcal{L_\text{penalty}}$ is of the same order of magnitude as the other likelihood terms, in order not to make it drive the reconstruction over the geometrical conditions, Eqs.~(\ref{eq:deltars},\ref{eq:deltard},\ref{eq:deltareq}). Overall, the total likelihood reads:
\begin{equation}
    \mathcal{L} = \mathcal{L}_s \times \mathcal{L}_d \times \mathcal{L}_\text{eq} \times \mathcal{L}_\text{penalty}\,,
    \label{eq:likelihood}
\end{equation}
where $\mathcal{L}_X$, with $X = \left\{s,d,\text{eq}\right\}$ are Gaussian likelihoods that implement conditions Eqs.~(\ref{eq:deltars},\ref{eq:deltard}) and (\ref{eq:deltareq}). More specifically, for the conditions Eqs.~(\ref{eq:deltars},\ref{eq:deltard}) I have:
\begin{equation}
    \mathcal{L}_{s,d} \propto \exp\left\{{-\frac{(\delta r_{s,d}^\star/r_{s,d}^\star - \text{target})^2}{2\sigma_{s,d}^2}}\right\}
\end{equation}
with $\text{target} = -0.07$ and $\sigma_{s,d} = 0.005$. The choice of $\sigma_{s,d}$ reflects the overall uncertainty in the high-and low-redshift measurements of $H_0$. On the contrary, since it is not an integral condition like Eqs.~(\ref{eq:deltars},\ref{eq:deltard}), I slightly relax the requirement for $\delta r_\text{eq}/r_\text{eq}$, and hence the reconstructed $\delta H/H_\text{fid}(z_\text{eq})$, Eq.~(\ref{eq:deltareq}), by allowing for a potential, rather generous, $\delta z_\text{eq}/z_\text{eq}\simeq 2\%$, in Eq.~(\ref{eq:condition_theta_eq}), and hence $\sigma_\text{eq} = 0.02$. 

Finally notice that an enhancement of the Hubble parameter in the decade of redshift preceding recombination can shift the redshift of last scattering, $z_\star$, toward higher values. Since $z_\star$ serves as the lower integration limit for Eqs.~(\ref{eq:deltars},\ref{eq:deltard}), this shift could, in principle, influence the reconstruction. However, this remains a subdominant effect since the conformal time of recombination is mainly determined by very well understood atomic physics, and it is therefore neglected in this analysis \cite{Knox:2019rjx}.

To practically reconstruct the $\delta H/H_\text{fid}(z)$, I parameterize it using $N_\text{knots} = 20$ control nodes distributed across the redshift range $z \in (1090, 20000)$. The reconstructed function is obtained by interpolating with a cubic spline the amplitude of the knots, which are varied to sample the likelihood in Eq.~(\ref{eq:likelihood}) via an MCMC algorithm. Chain convergence is monitored using the Gelman-Rubin $R - 1$ parameter \cite{Gelman:1992zz}, and chains are deemed converged when $R - 1 < 0.03$. Further details are provided in the \textit{Supplementary Material}.

\begin{figure*}[]
    \centering
    \includegraphics[width=0.8\linewidth]{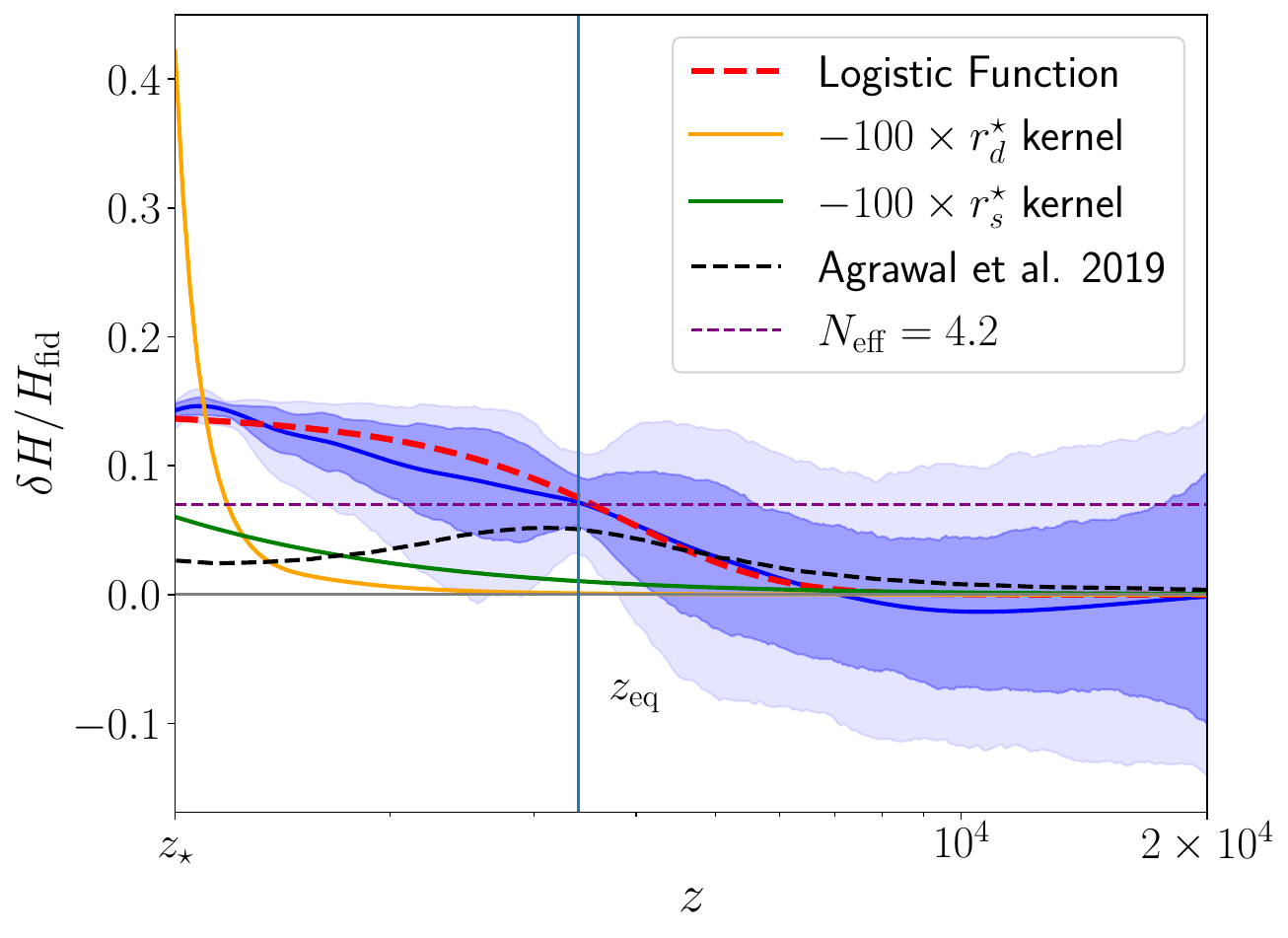}
    \caption{Reconstructed posterior for the fractional variation of the Hubble parameter $\delta H / H_\text{fid}$ relative to the reference $\Lambda$CDM model, obtained through cubic spline interpolation. The blue light bands indicate the 68\% and 95 \% credible intervals for the reconstructed function, while the solid blue line corresponds to the reconstructed mean function. Overlaid on the reconstructed posteriors, both the $(-100\times)r_s^\star$ (solid green) and $(-100\times)r_d^\star$ (solid yellow) kernels are represented, together with the proposed phenomenological logistic functions (dashed red and dashed purple). Large contours at high-redshift $(z > 4000)$, averaging to zero, are due to the low constraining power of CMB kernels $K_s$ and $K_d$ at very early times. Finally, the vertical line corresponds to the redshift of the fiducial matter-radiation equality, $z_\text{eq}$, while the dashed black and dashed purple lines correspond respectively to $\delta H/H_\text{fid}(z)$ of the bestfit to the $\phi^4$ Early-Dark Energy model \cite{Agrawal:2018own} and of the $N\text{eff} = 4.2$ model, as reported in \cite{Knox:2019rjx}.}
    \label{fig:deltaH}
\end{figure*}

\textbf{\textit{Results.}} --- The non-parametric reconstruction of the fractional deviation from the fiducial expansion history, $\delta H/H_\text{fid}(z)$, is shown in Fig.~\ref{fig:deltaH}. The figure displays the $68\%$ and $95\%$ credible intervals of the reconstructed posterior alongside the mean (solid blue line). I find a pronounced enhancement in $\delta H/H_\text{fid}$ starting around matter-radiation equality ($z \lesssim z_\text{eq}$) and reaching its maximum near recombination. At higher redshifts, the reconstructed function smoothly transitions into a long tail that averages to zero. To clarify the geometric origin of these features, Fig.~\ref{fig:deltaH} overlays the kernels for both the sound horizon at recombination, $K_s(z)$ (solid green), and the Silk damping scale, $K_d(z)$ (solid yellow), as well as the fiducial redshift of matter-radiation equality, $z_\text{eq}$ (vertical blue line). Crucially, the geometric properties of this transition, including its amplitude near recombination and the smooth decay at $z \gtrsim 4000$, are not arbitrary. Instead, they are mathematically determined by the overlaid kernels and the boundary condition on $\delta H/H_\text{fid}$ at matter-radiation equality. Because the Silk damping kernel $K_d(z)$ has narrow support and vanishes at high redshifts ($z \gtrsim 4000$), any modification to $H(z)$ introduced too early fails to influence the diffusion scale. Consequently, achieving a full $\simeq 7\%$ reduction in $r_d^\star$ requires the transition to reach an amplitude of $\delta H/H_\text{fid} \simeq 15\%$ localized within this narrow, pre-recombination window, see Fig.\ref{fig:deltaH}.\footnote{The interested reader can find in the \textit{Supplementary Material} a discussion on the validity of the linear regime for Eqs.~(\ref{eq:deltars},\ref{eq:deltard})).} The redshift range of the smooth transition ($z \simeq z_\text{eq}$) is dictated by an interplay between the sound horizon kernel $K_s(z)$, which favors a drop at $z \lesssim z_\text{eq}$ to limit high-redshift contributions, and the condition on $\delta H/H_\text{fid}$ at matter-radiation equality in Eq.~(\ref{eq:deltareq}), which forces the reconstruction to drop at $z \gtrsim z_\text{eq}$.

To gain analytical intuition, I parametrize the reconstructed $\delta H/H_\text{fid}$ using a logistic function:
\begin{equation}
    \delta H/H_\text{fid}(z) = \frac{\delta_\text{max}}{1 + e^{(z-z^\dagger)/\sigma}},
\end{equation}
defined by an amplitude $\delta_\text{max} = 0.15$, a transition redshift $z^\dagger = z_\text{eq}$, and a width $\sigma = 1000$ (dashed red curve). By achieving the target $7\%$ reduction within 1 $\sigma$, in both $r_s^\star$ and $r_d^\star$ (see Tab.~\ref{tab:table}), this minimal three-parameter model emphasizes the essential features of the reconstruction: the highly constrained amplitude immediately before recombination, evident also in the thinning of the confidence intervals, and the long, zero-average tail at $z \gtrsim 4000$. However, both this parametrization and the mean reconstructed function slightly under-reduce $r_d^\star$. This occurs because both amplitudes are slightly smaller than the value required to achieve a full $7\%$ reduction in the Silk damping scale, immediately prior to recombination. On the other hand, as demonstrated by the logistic model, larger amplitudes before matter-radiation equality would inevitably lead to an excessive $\delta r_s^\star/r_s^\star$. Indeed, $\delta H/H_\text{fid}$ cannot behave as a simple step function to satisfy the $\delta r_s^\star/r_s^\star$ and $\delta r_d^\star/r_d^\star$ integrals, Eqs.~(\ref{eq:deltars},\ref{eq:deltard}), and then drop immediately before $z_\text{eq}$; it must also fulfill the condition on $\delta H/H_\text{fid}(z_\text{eq}) \simeq 0.07$, Eq.~(\ref{eq:deltareq}), therefore decaying much slower. Because of the long tail of $K_s(z)$, in its slow decay, $\delta H/H_\text{fid}$, integrates contributions to the $\delta r_s^\star/r_s^\star$ integral also at high redshifts, $z > z_\text{eq}$, severely constraining higher amplitudes of $\delta H/H_\text{fid}$ closer to recombination. This \textit{tug-of-war} results in a reconstruction that slightly under-reduces the Silk damping scale. This feature underscores both the high sensitivity and the very narrow support of the damping scale kernel $K_d(z)$, which poses stringent background constraints on the amplitude of deviations from the fiducial cosmology around recombination.

\begin{table}[]
\centering
\renewcommand{\arraystretch}{1.5} 
\scalebox{0.8}{
\begin{tabular}{?c?c|c|c|c|c?}
\thickhline

\multicolumn{1}{?c?}{\textbf{Curves}} & \textbf{$  \delta r_s^\star/r_s^\star$} & \textbf{$\delta r_d^\star/r_d^\star$} & $\% \,\delta r_s^\star/r_s^\star$ ($z < z_{\text{eq}}$) & $ \% \,\delta r_d^\star/r_d^\star$ ($z < z_{\text{eq}}$) \\
\thickhline
Mean 
    & $-0.0702$ & $-0.0672$ & $91.4 \, \%$ & $99.4 \, \%$  \\
\hline
Logistic 
    & $-0.0751$ & $-0.0651$ & $90.5 \, \%$ & $99.3 \, \%$ \\
\thickhline
\end{tabular}}
\caption{Fractional reduction of the sound horizon ($r_s^\star$) and Silk damping ($r_d^\star$) scales at recombination, along with their percentage contributions from before the fiducial matter-radiation equality ($z < z_{\text{eq}}$).}
\label{tab:table}
\end{table}

\textbf{\textit{Discussion.}} --- The exquisite precision of CMB measurements dictates that any early-universe resolution to the Hubble tension must strictly preserve the observed ratios of three angular scales: $\theta_s^\star$, $\theta_d^\star$, and $\theta_\text{eq}$. The importance of these angular scales was first discussed in \cite{Hu:1996mn,Hu:2000ti,Hu:2001bc} and later emphasized in the context of the Hubble tension \cite{Knox:2019rjx,Poulin:2018cxd}, where it was noted that the fit to these scales provides a \textit{necessary} condition but does not guarantee an overall good fit to the full CMB spectrum \cite{Adil:2022hkj}. More specifically, the latter is due to the inability of $\theta_\text{eq}$ to fully capture the radiation driving envelope of the CMB. 

That being said, any viable pre-recombination modification to the expansion history must induce equivalent fractional shifts in all three scales, such that $\delta r_s^\star/r_s^\star\simeq \delta r_d^\star/r_d^\star \simeq \delta r_\text{eq}/r_\text{eq}$. However, because the damping scale kernel $K_d(z)$ strongly decays at high redshifts ($z \gtrsim 4000$) while the sound horizon scale kernel $K_s(z)$ remains broadly active, extended or high-redshift injections of energy---such as those characteristic of extra-relativistic degrees of freedom and standard Early Dark Energy models---fall short on this geometric requirement, by inevitably decoupling the $r_s^\star$ and $r_d^\star$ scales. As demonstrated by my model-independent reconstruction, the expansion rate is instead mathematically forced into a smooth transition, happening around the redshift of fiducial matter-radiation equality and reaching its maximum immediately before recombination. The functional shape of the transition identified here is not an artifact of a specific amplitude, but rather a universal feature. Whether a micro-physical model aims for a full $\simeq 15\%$ enhancement to attempt to entirely resolve the tension, or a smaller shift as part of a hybrid cosmology, featuring departures from $\Lambda$CDM both pre- and post-recombination, \textit{any successful early-universe mechanism must inherently mimic the reconstructed behavior}. As argued above, the localized sensitivity of the Silk damping kernel $K_d(z)$, imposes a stringent \textit{lower bound} on the amplitude of the transition. Indeed, achieving the necessary $\simeq 7\%$ reduction in $r_d^\star$ strictly anchors the requisite amplitude of the reconstructed $\delta H / H_\text{fid}$ to $ \simeq 15\%$ within the narrow redshift window where the $r_d^\star$ kernel is active. Concurrently, the duration of the transition is constrained by the interplay between the sound horizon kernel $K_s(z)$ and the condition on $\delta r_\text{eq}/r_\text{eq}$, forcing $\delta H/H_\text{fid}(z_\text{eq} )\simeq 7 \%$, as discussed above. Consequently, the background kinematics permits little theoretical freedom: a complete early-universe resolution intrinsically mandates this significant $\simeq 15\%$ enhancement, immediately before recombination. Regardless of the hypothetical physical scenario provoking it, such a drastic shift could \textit{potentially} generate prohibitive perturbation-level artifacts in the CMB power spectrum, such as an excessive Early Integrated Sachs-Wolfe (eISW) effect \cite{Vagnozzi:2021gjh}. I remain agnostic whether such a transition could fit the full CMB power spectrum, and I leave the study of perturbation-level effects to future investigations. However, the discussed geometric rigidity of the reconstructed expansion history $\delta H/ H_\text{fid}$ could effectively \textit{hint} to a no-go theorem for purely pre-recombination solutions to the Hubble tension \cite{Vagnozzi:2023nrq,Toda:2024ncp}. 

Despite being less evident, the need for a combined pre-and post-recombination mechanism is suggested by the reconstructed expansion history already at the background level. Indeed, taken at face value, and ignoring for the moment potential issues related to perturbation-level effects, Fig.~\ref{fig:deltaH} shows that the geometric CMB constraints mandate $\delta H/H_\text{fid}(z_\star) \simeq 15\%$. If this shift remained constant until today, it would produce an equivalent boost in $H_0$, over-enhancing the Hubble constant and destroying the fit to the angular scales. In fact, if the post-recombination expansion history is unaltered, as dictated by purely pre-recombination proposals, a much smaller shift of $\delta D_A^\star/D_A^\star \simeq -\delta H_0/H_{0,\text{fid}} \simeq -7\%$ is required to keep the angular scales from changing when the scales Eqs.~(\ref{eq:deltars},\ref{eq:deltard},\ref{eq:deltareq}) are changed. This creates a clear incompatibility with the $15\%$ shift of $\delta H/H_\text{fid}$, required at $z_\star$. Hence new, post-recombination, dynamics in $\delta H/H_\text{fid}$ is needed to compensate for the would-be \text{over}-enhancement of $H_0$. As a consequence, one can postulate a second transition, happening during the dark ages and being effectively \textit{invisible} to current \textit{background} cosmological probes; a visual example is provided in the \textit{Supplementary material}. However, such a transition, despite looking innocuous at the background level, poses some serious challenges once perturbations are also taken into account. Indeed, a $\simeq 8\%$ $\delta H/H_\text{fid}$ transition is expected to produce visible effects in the CMB power spectrum via \textit{mid}ISW effect, potentially constraining this possibility. 

It is worth noting that some authors have argued this geometric rigidity could be partially relaxed if the enhanced diffusion damping is compensated by a concurrent increase in the primordial scalar spectral index, $n_s$, alongside the physical matter density, $\omega_m$ \cite{Knox:2019rjx, Poulin:2018cxd, Agrawal:2019lmo, Smith:2025zsg}. Nevertheless, this parameter shift introduces its own observable signatures: the increased $n_s$ and matter density amplify small-scale matter clustering, which in turn leads to stronger CMB lensing. This lensing contribution creates a noticeable enhancement of power at high multipoles ($\ell \gtrsim 3000$), meaning the damping decrement and $n_s$ effects do not perfectly cancel and can instead be robustly disentangled by high-resolution, small-scale CMB observations \cite{Smith:2025zsg}. Another worth mentioning scenario, which could potentially escape the abovementioned geometric rigidity, is realized by varying electron mass mechanisms. The latter, by raising the binding energy of neutral Hydrogen, induces early recombination \cite{Sekiguchi:2020teg, Hart:2019dxi,Lee:2022gzh}, therefore reducing the sound horizon scale at recombination $r_s^\star$. The scaling $\sigma_T \propto m_e^{-2}$ helps maintaining the $\theta_s^\star/\theta_d^\star$ ratio, as the integral for $r_d^\star$ scales as $1/\sqrt{\sigma_T}$, see Eqs.~(\ref{eq:rX},\ref{eq:F_rd}). This allows for additional contributions to $\delta r_d^\star/r_d^\star$ to balance the reduction in the sound horizon scale at recombination $r_s^\star$ \cite{Hart:2019dxi, Hart:2017ndk}. Accounting for this scaling opens a large geometrical degeneracy between $m_e$ and $H_0$. When combined with a non-zero spatial curvature ($\Omega_K < 0$), this mechanism can provide a higher $H_0$ and increased $m_e$ while maintaining a good fit to CMB, BAO, and SNeIa data \cite{Sekiguchi:2020teg}.

Beyond purely background-level consideration, I leave to future work the integration of this phenomenological step into full Boltzmann codes (such as CAMB \cite{Lewis:1999bs} or CLASS \cite{Blas:2011rf}) to directly confront the CMB angular power spectra, both using specific fluid or scalar field prescriptions, and via model-independent reconstructions incorporating generalized perturbation dynamics.  Furthermore, the smooth background transition identified here serves as a stringent geometric blueprint to inspire fundamental model building, particularly for mechanisms driving early-universe transitions. If pre-recombination modifications alone prove insufficient to fully resolve the Hubble tension, this framework naturally motivates the exploration of hybrid cosmologies that combine this early-time background step with post-recombination new physics. In such interconnected scenarios, my approach could be extended to calculate a robust, model-independent upper bound on $H_0$, from late-time reconstructions of the expansion history of the Universe, using unanchored supernovae to constrain the remaining viable parameter space.

\textbf{\textit{Acknowledgments.}} --- I am particularly grateful to Sunny Vagnozzi and William Giarè for precious discussions. I acknowledge support from the Istituto Nazionale di Fisica Nucleare (INFN) through the Commissione Scientifica Nazionale 4 (CSN4) Iniziativa Specifica “Quantum Fields in Gravity, Cosmology and BHs” (FLAG). This publication is based upon work from the COST Action CA21136 “Addressing observational tensions in cosmology with systematics and fundamental physics” (CosmoVerse), supported by COST (European Cooperation in
Science and Technology).

\bibliography{hubble}
\clearpage
\onecolumngrid
\newpage

\section*{Supplementary Material}

\section{Condition on $\delta r_\text{eq}/r_\text{eq}$}
Starting from the definition of particle horizon at matter radiation equality:
\begin{equation}
    r_\text{eq} \equiv \frac{1}{a_\text{eq}H(a_\text{eq})} = \frac{1+z_\text{eq}}{H(z_\text{eq})} \simeq\frac{z_\text{eq}}{H(z_\text{eq})},
\end{equation}
where $z_\text{eq}$ is the redshift of matter-radiation equality of the fiducial model, and by assuming that deviations $\delta H/H_\text{fid}$ from the fiducial $H_\text{fid}$ could potentially cause shifts in $z_\text{eq}$, such that $z_\text{eq}^{\text{new}} = z_\text{eq} + \delta z_\text{eq}$, and $r_\text{eq}^\text{new} = r_\text{eq} + \delta r_\text{eq}$, I find:
\begin{equation}
    \begin{split}
        \frac{\delta r_\text{eq}}{r_\text{eq}} &= \frac{1}{r_\text{eq}}\left[r_\text{eq}^\text{new} - r_\text{eq}\right]\\
        &= \frac{1}{\frac{z_\text{eq}}{H(z_\text{eq})}} \left[\frac{z_\text{eq} + \delta z_\text{eq}}{H(z_\text{eq}) + \delta H(z_\text{eq})} - \frac{z_\text{eq}}{H(z_\text{eq})}\right]\\
        & =  \frac{1}{\frac{z_\text{eq}}{H(z_\text{eq})}}\left[\frac{z_\text{eq} + \delta z_\text{eq}}{H(z_\text{eq})}\left(1 - \frac{\delta H}{H}(z_\text{eq})\right) - \frac{z_\text{eq}}{H(z_\text{eq})}\right].
    \end{split}
\end{equation}
Simplifying the above expression and dropping second-order terms, I find  Eq.~(\ref{eq:condition_theta_eq}).
%
\section{Roubustness check of the linear approximation in Eqs.~(\ref{eq:deltars},\ref{eq:deltard})}
As shown in text, the geometric conditions imposed by the CMB angular scales mandate a $\simeq 15\%$ $\delta H/H_\text{fid}$ amplitude around recombination. A quick calculation shows that, for such an amplitude the first order approximation
\begin{equation}
    \frac{1}{1+\delta H/H_\text{fid}} \simeq 1-\delta H/H_\text{fid} 
    \label{eq:first_order}
\end{equation}
is only precise roughly at the $\simeq 2\%$ level. However it is important to note that, while the kernels are most sensitive around recombination, the overall shifts $\delta r_X/r_X$ are defined by integral conditions. As a consequence the shifts $\delta r_X/r_X$ integrate contributions across the entire the redshift domain, where the reconstructed $\delta H/H_\text{fid}$ remains mostly well within the linear regime approximation. Furthermore, the linear approximation of Eq.~(\ref{eq:first_order}), \textit{underestimates} the contribution of $\delta H/H_\text{fid}(z\simeq z_\star)$ to Eqs.~(\ref{eq:deltars},\ref{eq:deltard}), therefore leading to a \textit{conservative} estimate of the reconstructed $\delta H/H_\text{fid}(z \simeq z_\star)$. To validate these arguments, I compare the shifts $\delta r_s^\star/r_s^\star$ and $\delta r_d^\star/r_d^\star$ calculated using the linear approximation, i.e., Eqs.~(\ref{eq:deltars},\ref{eq:deltard}), against those derived from the exact formula. In particular, I use the mean reconstructed function to compute the shifts using the first order formula:
\begin{equation}
     \frac{\delta r_X}{r_X} = \int_{z_\star}^\infty K_X(z)\left[\frac{\delta H}{H_\text{fid}}(z)\right]_\text{mean},
\end{equation}
and the exact formula:
\begin{equation}
    \frac{\delta r_X}{r_X} = \int_{z_\star}^\infty K_X(z)\frac{\left[\delta H/H_\text{fid}(z)\right]_\text{mean}}{1 + \left[\delta H/H_\text{fid}(z)\right]_\text{mean}}.
\end{equation}
As expected, the exact formula produces larger shifts in both $r_s^\star$ and $r_d^\star$, by $0.03\%$ and $0.17\%$ respectively. Given the chosen $\sigma_X$, see both the main text and the \textit{Likelihood} subsection of \textit{Code and Algorithm details}, these effects are completely negligible. The error in $\delta r_d^\star/r_d^\star$ is an order of magnitude larger than that of $\delta r_s^\star/r_s^\star$ because the photon diffusion kernel, $K_d(z)$, integrates the majority of its contributions in the region where $\delta H/H_\text{fid}(z) \simeq 15\%$. Finally Eq.~(\ref{eq:deltareq}) was not considered in this discussion as, by construction, around $z_\text{eq}$, $\delta H/H_\text{fid} \simeq 0.07$, which is well within the linear regime.

\section{Post-recombination transition to guarantee consistency}
\begin{figure}[!h]
    \centering
    \includegraphics[width=0.6\linewidth]{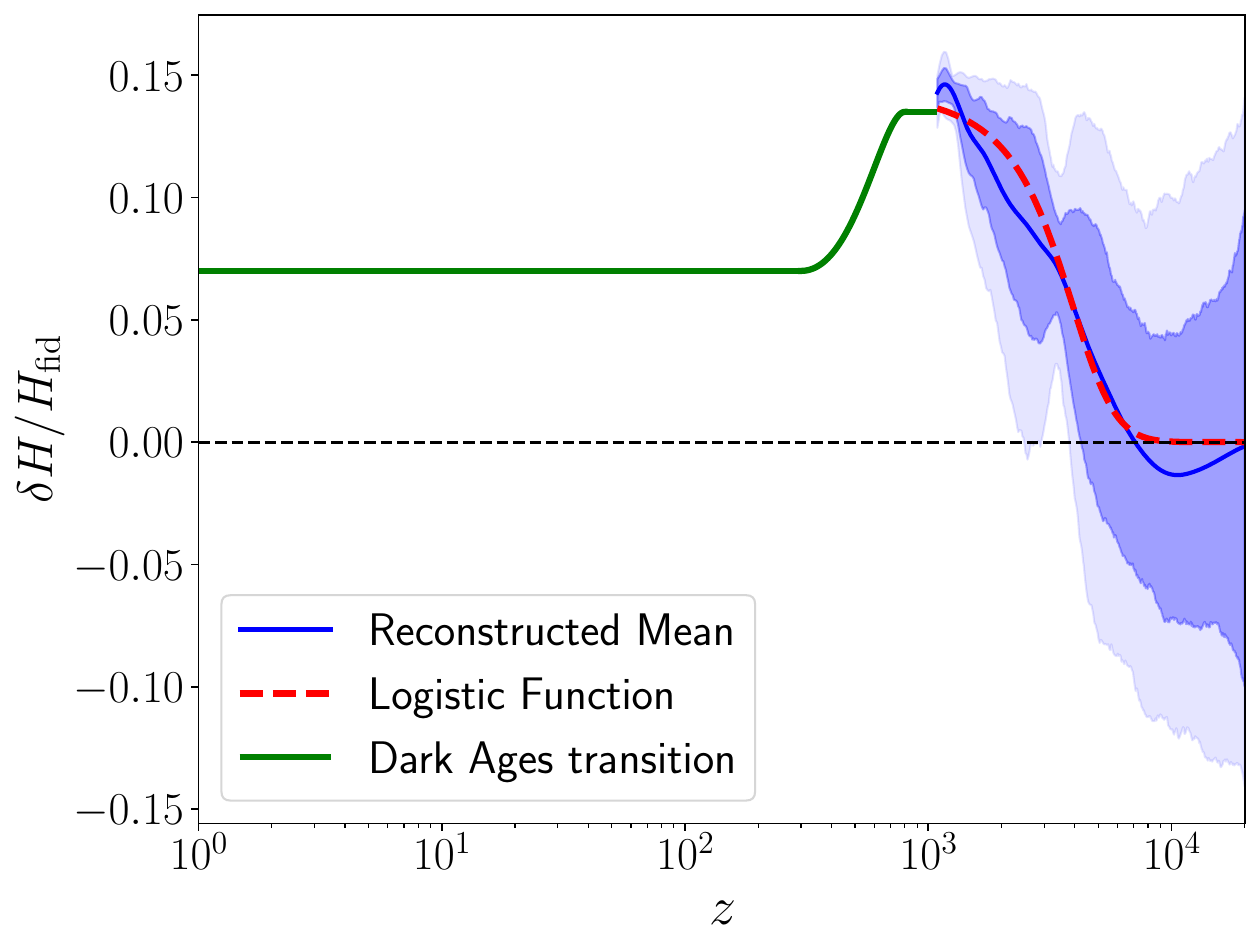}
    \caption{Background transition happening during the dark ages to reconcile the over-enhanced $\delta H/H$ found by the model-independent reconstruction with the low-redshift measurements of $H_0$.}
    \label{fig:invisible_transition}
\end{figure}

Taken at face value, the reconstructed function of Fig.~\ref{fig:deltaH} suggests the need for a second, post-recombination, transition in $\delta H/H_\text{fid}(z)$, to correct for the would-be over-enhanced $H_0$. In order to be invisible to background low-redshift probes, as well as to maintain the fit to the CMB angular scales, such a transition would need to happen at sufficiently high redshift. Indeed, for a given post-recombination transition to be acceptable, at least at the background level, the condition that $\delta D_A^\star/D_A^\star \simeq -\delta H_0/H_\text{0,fid} \simeq -7\%$ must hold, in order to compensate for the equivalent shifts in the comoving scales $\delta r_s^\star/r_s^\star \simeq \delta r_d^\star/r_d^\star \simeq \delta r_\text{eq}/r_\text{eq}$. Since the high-redshift contributions to the angular diameter distance integral are suppressed by $1/E(z)$:
\begin{equation}
    D_A^\star = \int_0^{z_\star}\frac{dz}{H(z)} = \int_0^{\simeq 400}\frac{dz}{H(z)} + \cancelto{\lesssim 2\%}{\int_{400}^{z_\star}\frac{dz} {H(z)}} \simeq \int_0^{\simeq 400}\frac{dz}{H(z)},
\end{equation}
a transition happening at sufficiently high redshifts, i.e., deep in the dark ages, would be effectively \textit{invisible} to $D_A^\star$, and therefore it would simultaneously be able to compensate for the would-be over-enhanced $H_0$ and to maintain the fit to the CMB angular scales. As proof-of-principle and without aiming for generality, Fig.~\ref{fig:invisible_transition} shows a plausible, dark ages transition (solid green), smoothly interpolating between the reconstructed function and the value of $\delta H/H_\text{fid}(z=0) \simeq 7\%$ inferred from low-redshift measurements of the Hubble constant $H_0$ \cite{Riess:2021jrx}. Specifically, the transition showed in Fig.~\ref{fig:invisible_transition} provides $\delta D_A^\star/D_A^\star \simeq -7\%$ with a precision of:
\begin{equation}
    \Delta \equiv\frac{\left(\delta D_{A}^\star/D_A^\star\right)_\text{transition} - \left(\delta D_{A}^\star/D_A^\star\right)_\text{no,transition}}{\left(\delta D_{A}^\star/D_A^\star\right)_\text{no,transition}} \simeq 0.3 \%
\end{equation}
where $\left(\delta D_{A}^\star/D_A^\star\right)_\text{no,transition} \simeq 7\%$, therefore fulfilling the background constraints. However, as stressed in the text, dark ages transition are severely constrained by perturbation-level effects such as \textit{mid}ISW. 

\section{Code and algorithm details}
I describe here the algorithm implemented to non-parametrically reconstruct the pre-recombination expansion history of the Universe. As mentioned above, I performed the reconstruction using a Markov chain Monte Carlo (MCMC) method to sample the parameter space given by the knot heights of the reconstructed function. 

\subsection{Initialization}
$N_\text{knots} = 20$ knots were used to reconstruct the function $\delta H/H(z)$ via cubic spline interpolation, across the redshift range $z \in (z_\star,20000)$, where $z_\star = 1090$ \cite{Planck:2018vyg}.  The algorithm proved robust to changes in the number of knots, and $N_\text{knots} = 20$ was chosen as a good compromise between resolution power and convergence time. Knowing that the sensitivity of the sound horizon and damping horizon kernels $K_s(z)$ and $K_d(z)$ is mostly confined close to recombination, the knots were chosen in such a way to provide higher resolution power around this region. More specifically, the knots were subdivided into three regions: a \textit{dense} region, $z \in (z_\star,3800)$, containing 50$\%$ of $N_\text{knots}$, a \textit{medium populated} region, $z \in (3800,10000)$, containing $30\%$ of $N_\text{knots}$ and finally a \textit{sparse} region, $z \in (10000,20000)$, containing $20\%$ of $N_\text{knots}$. The knot heights were initialized randomly from a flat distribution $\mathcal{U}[-0.1,0.1]$. Anyway, the algorithm proved robust to changes in the width of the initial knot heights flat distribution. 

\subsection{Priors}
\begin{figure}[!t]
    \centering
    \includegraphics[width=0.6\linewidth]{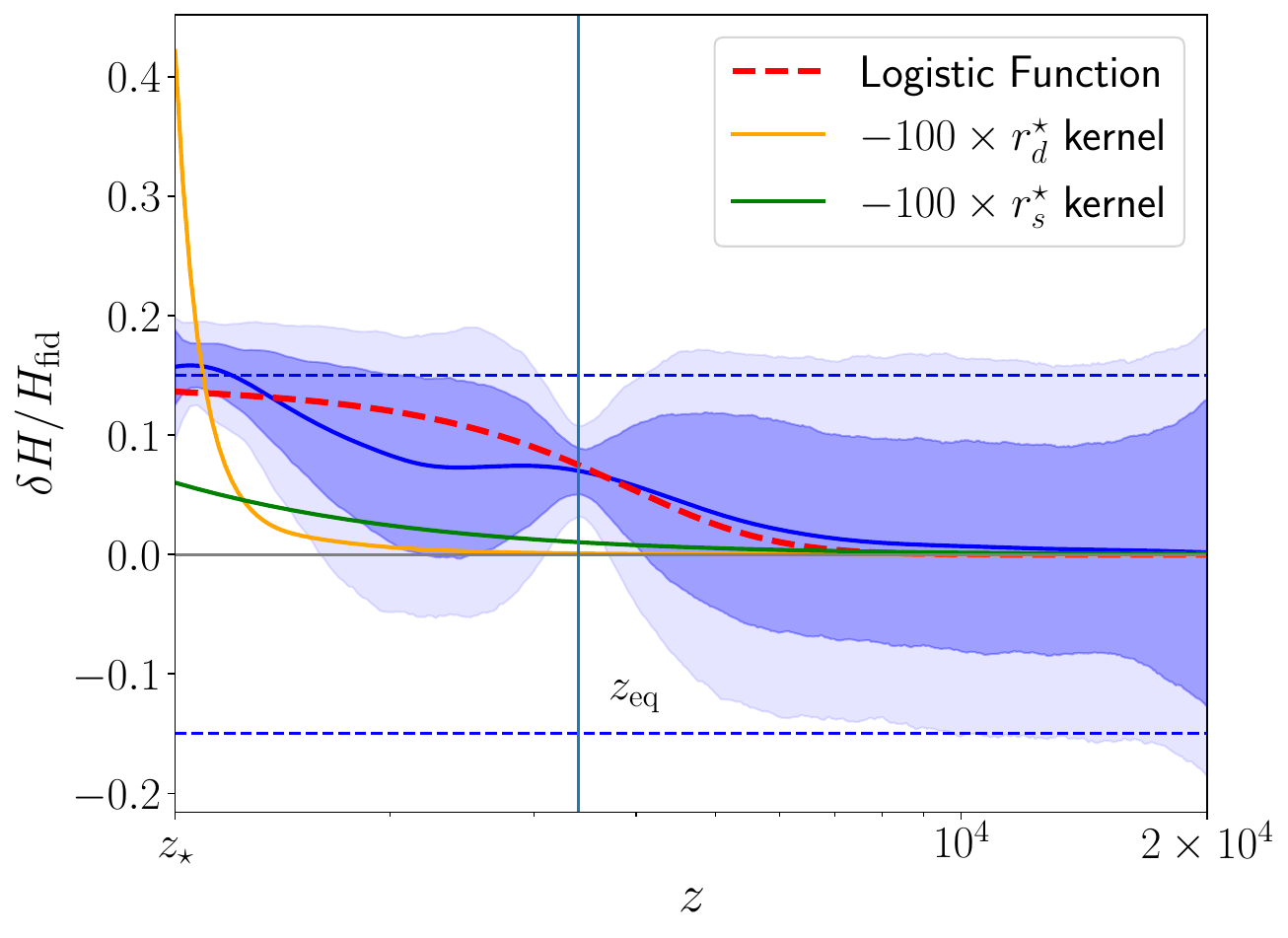}
    \caption{Same as Fig.~\ref{fig:deltaH}, but with a larger prior $|\delta H/H_\text{fid}(z_i)| \leq 0.2$. The reconstructed function basically reproduces all the features discussed above. The two horizontal dashed blue lines indicate the prior region of the reconstructed function of Fig.~\ref{fig:deltaH}.}
    \label{fig:prior_02}
\end{figure}

I adopted the following priors for the parameters
\begin{equation}
    \Big|\delta H/H_\text{fid}(z_i)\Big| \leq 0.15, \qquad z_i \in \left\{z_1,z_2,...,z_{N_\text{knots}}\right\},
    \label{eq:prior}
\end{equation}
which are motivated by the linear-response approach, upon which the whole framework is developed, see Eqs.~(\ref{eq:deltars},\ref{eq:deltard}, \ref{eq:deltareq}). The exact threshold where non-linear effects start to become relevant is hard to discriminate. However, I chose the rather conservative bound of Eq.~(\ref{eq:prior}), based on two arguments: \textit{first}: by trial and errors the prior choice of Eq.~(\ref{eq:prior}) proved enough to achieve the target reductions in both $r_s^\star$ and $r_d^\star$, see Tab.~\ref{tab:table}, while having the quadratic corrections $\mathcal{O}((\delta H/H_\text{fid})^2)$ comfortably negligible as compared to the linear response, and \textit{second}: runs with larger prior bounds reproduce essentially the same features as those observed above; see for example Fig.~\ref{fig:prior_02}. Of course the latter case does not hold for choices of smaller prior bounds, e.g. $|\delta H/H_\text{fid}| \leq 0.1$ since, as was stressed in the text, the geometric rigidity of the kernels, mandate a $\simeq15\%$ amplitude of $\delta H/H(z)$ immediately before recombination. Finally, it is worth noticing that due to the nature of cubic spline interpolations, this prior does not prevent the reconstructed function from wiggling around and eventually exceeding the hard prior bound between subsequent points. A more stringent constraint would be to exponentially penalize the reconstructed function exceeding the linear regime threshold. However, due to the tug-of-war between the various kernels described above, such high amplitude wiggles are already disfavored, with the only exception of the redshift region $z \in (z_\star,\simeq 1200)$ immediately prior to recombination.

\subsection{Likelihoods}
As described above, I adopt Gaussian likelihoods of the form:
\begin{equation}
    \mathcal{L}_X \propto\exp\left\{-\frac{(\delta r_X/r_X - \text{target}_X)^2}{2\sigma_X^2}\right\},
    \label{eq:like}
\end{equation}
where $X = \left\{s,d\right\}$, and the choices of both targets and $\sigma$s are \textit{guided} by observations. More specifically, I choose $\delta r_s^\star/r_s^\star\simeq -7\%$, and as a consequence $\delta r_d^\star/r_d^\star \simeq -7\%$, based on considerations like:
\begin{equation}
    r_s^\star H_0 \simeq \text{const} \qquad\text{\&}\qquad \frac{\delta H_0}{H_0}  = \frac{H_0^{\text{low-redshift}} - H_0^{\text{high-redshift}}}{H_0^{\text{high-redshift}}} \simeq 7\% ,
\end{equation}
where \textit{low}- and \textit{high}-redshift are broadly referred to local-universe \cite{Riess:2021jrx,H0DN:2025lyy} and early-universe \cite{Planck:2018vyg,Schoneberg:2019wmt} measurements of $H_0$, respectively. The choice of $\sigma_X$ for $X = \left\{s,d\right\}$ is dictated by the scatter in the central values of the Hubble constant, as measured by different probes. For this reason, I choose a rather generous $\sigma_{r,s} = 0.005$. On the other hand, the condition on $\delta r_\text{eq}/r_\text{eq}$, Eq.~(\ref{eq:deltareq}), is \textit{mathematically} different from the others, as it is not an integral constraint, but rather a condition on $\delta H/H_\text{fid}$ at a given redshift; i.e., at $z=z_\text{eq}$. More specifically, by requiring $\delta H/H_\text{fid}(z_\text{eq}) \simeq 0.07$, Eq.~(\ref{eq:condition_theta_eq}) provides a \textit{necessary} condition for having both $\delta z_\text{eq}/z_\text{eq} \simeq 0$ and  $\theta_\text{eq} \simeq \text{const}$. For this reason, the variable in the gaussian likelihood, Eq.~(\ref{eq:like}), is $\delta H/H_\text{fid}(z_\text{eq})$, with $\text{target}_\text{eq} = 0.07$. For the latter I chose a rather conservative $\sigma_\text{eq} = 0.02$, which reflect a certain degree of tolerance of $\delta z_\text{eq}/z_\text{eq}$ in the CMB, found in some known pre-recombination scenarios \cite{Poulin:2023lkg}. Finally, a penalty term Eq.~(\ref{eq:pen_like}) is included to suppress high-frequency oscillations in the reconstruction. To prevent the penalty from driving the fit, the parameter $\lambda$ was calibrated so that the resulting likelihood contribution remains comparable in magnitude to the physical contributions. Through trial and error, $\lambda = 10^{-4}$ was found to satisfy this condition. Notably, the reconstructed posteriors are robust to variations in $\lambda$ within approximately one order of magnitude around this value.

\subsection{Burn-in, adaptive logic \& convergence}
I implemented a burn-in phase of $N=10^5$ steps, determined through trial and error, preceding the main sampling. To speed up convergence, I chose initial step sizes that reflect the sensitivity of the kernels: $\mathcal{S}_\text{low} = 0.001$ for $z_i < z_\text{eq}$ and $\mathcal{S}_\text{high} = 0.005$ for $z_i > z_\text{eq}$. Furthermore, I employed an adaptive mechanism to adjust the proposal width toward a target acceptance rate of $\sim 20\%$. To guarantee ergodicity, four chains were independently initialized, and convergence was established when the Gelman-Rubin parameter reached $R - 1 < 0.03$ \cite{Gelman:1992zz}.

\end{document}